**Modeling Policy and Resource Dynamics in the Construction Sector of Developing Countries: A System Dynamics Approach Using Sudan as a Case Study**

**Malik Dongla, PhD**
CEO
Malik Dongula for Construction & Contracting
Khartoum, Sudan
Email: malikdongula@malikdongula.com

**Mohamed Khalafalla, PhD, MBA, PMP**
Associate Professor
Construction Engineering Technology
Florida A&M University, Tallahassee, FL, 32307
Email: mohamed.ahmed@famu.edu
ORCID: 0000-0001-9621-4940

**ABSTRACT**
Construction industries in developing countries face systemic challenges such as chronic project delays, cost overruns, and regulatory inefficiencies. This paper presents a system dynamics (SD) modeling framework for analyzing policy and resource dynamics within Sudan's construction sector, with broader applicability to Least Developed Countries (LDCs). The model incorporates key variables related to workforce, material supply, financing, and policy delays, and is calibrated using genetic algorithms (GAs) based on sectoral data and expert input. Simulation results across four policy scenarios indicate that regulatory reform and workforce training are the most effective levers for improving project performance. Specifically, implementing streamlined regulatory procedures reduced project delays by up to 32%, while investment in human capital decreased cost overruns by 28% over a 10-year simulation horizon. In contrast, scenarios focusing solely on material supply or financial inputs produced limited gains without corresponding policy or labor improvements. Sensitivity analysis further revealed that the system is highly responsive to macroeconomic stability and public investment flows. The study demonstrates that a hybrid SD-GA modeling approach offers a valuable decision-support tool for policymakers seeking to improve infrastructure delivery under uncertainty. Recommendations include phased regulatory reforms, targeted capacity building, and integrating modeling tools into strategic infrastructure planning in LDCs.

**Keywords:** System Dynamics, Genetic Algorithms, Construction Industry, Developing Countries, Policy Simulation, Sudan, Infrastructure Planning

INTRODUCTION
The construction industry plays a vital role in driving economic growth, employment, and infrastructure development, particularly in developing countries (*1–3*). Yet, in many Least Developed Countries (LDCs), the sector is plagued by systemic inefficiencies, including chronic project delays, cost overruns, poor coordination, and fragmented policy implementation. These issues are often compounded by external factors such as inflation, political instability, and underinvestment in human capital (*4, 5*). Sudan offers a representative case of these challenges. Despite a high demand for infrastructure stemming from urban expansion, population growth, and post-conflict reconstruction needs, Sudan's construction industry suffers from frequent disruptions, weak regulatory enforcement, and limited resource planning (*6, 7*). The interdependencies between labor availability, material supply, project financing, and regulatory bottlenecks make traditional project management approaches insufficient. A dynamic, systems-based perspective is needed to uncover and address the feedback loops driving inefficiencies.

System Dynamics (SD) modeling offers a promising tool for tackling complex, feedback-driven systems such as the construction industry. Originally developed for industrial and organizational analysis, SD has been increasingly applied in infrastructure planning, resource management, and public policy modeling (*8, 9*), but its application in the construction sector of developing countries remains limited. Compounding the challenge is the scarcity of reliable data in many LDCs, which makes model calibration difficult. To overcome this, Genetic Algorithms (GAs) can be integrated with SD models to optimize parameter estimation and enhance model accuracy, even with limited datasets (*10, 11*). This paper presents a hybrid SD-GA modeling approach tailored to Sudan's construction sector. The objectives are threefold: [1] to model the interactions among policy, labor, and material subsystems that influence project delivery; [2] to simulate various policy intervention scenarios; and [3] to evaluate which strategies yield the greatest improvements in project performance. The model is designed to support infrastructure policymakers and planners in LDCs by offering insights into high-leverage interventions that enhance efficiency and reduce delays.

LITERATURE REVIEW
**System Dynamics in Construction**
SD has emerged as a powerful tool to model complex systems involving time delays, feedback loops, and nonlinear relationships, characteristics common in large-scale infrastructure and construction projects. Early applications of SD in construction management focused on modeling project delays, resource allocation, labor productivity, and learning curves (*12, 13*). For example, Park (*14*) used SD to simulate the effects of productivity changes on project timelines, while Ford and Sterman (*15*) examined rework cycles and the impact of schedule pressure on quality. Over time, SD has been applied to more strategic areas such as supply chain coordination, risk assessment, and lifecycle planning. Research by Alzahrani and Emsley (*16*) integrated SD into decision support systems for project risk analysis, while Ogunlana et al. (*17*) applied SD to evaluate long-term infrastructure development under uncertainty. These applications underscore SD's versatility in capturing real-world complexity and testing alternative scenarios for system improvement. However, most existing SD applications have been concentrated in high-income countries with stable institutional structures and readily available data. In these contexts, models are often tailored for high-rise building projects, mega-infrastructure ventures, or digital construction workflows (*18*).

**SD in Developing Countries**
In contrast, SD applications in construction industries of developing countries remain limited, both in number and depth. Where they exist, they often focus on isolated project-level phenomena rather than sector-wide systemic challenges. For instance, studies in Nigeria and India have used SD to assess material delays or workforce scheduling, but few have integrated policy, labor, supply, and macroeconomic variables into a unified sector model (*17, 19, 20*). Sudan's construction sector is emblematic of broader challenges in LDCs, ranging from unstable procurement systems and fluctuating exchange rates to severe labor shortages and fragmented institutional mandates **(Table 1)** (*21–23*). Yet, very little academic work exists that models these dynamics in a cohesive and quantitative way.

**TABLE 1. Summary of Key Challenges in Sudan's Construction Sector**

| Category | Example Challenges | Sources |
|---|---|---|
| Regulatory | Overlapping mandates, delayed permitting | (*23, 24*) |
| Workforce | Shortage of skilled labor, high turnover | (*25, 26*) |
| Materials Supply | Fluctuating import prices, transport bottlenecks | (*27, 28*) |
| Financial Access | Inflation, unstable credit systems | (*29, 30*) |
| Project Management | Inadequate planning, absence of feedback control | (*31, 32*) |

**Complementary Tools: Genetic Algorithms for Model Calibration**
One of the persistent challenges in applying SD in developing countries is the lack of granular time-series data needed for robust model calibration. To address this, several researchers have introduced optimization techniques, particularly GAs, to improve the calibration of SD models by automating parameter estimation through evolutionary search processes (*33–35*). GAs have been used to fine-tune complex models where traditional estimation methods fall short, particularly in cases involving fuzzy data or qualitative judgments. For example, Sohrabinejad et al. (*36*) successfully applied GAs to calibrate SD models in broadcasting policy settings, while Akopov et al. (*37*) applied GAs to simulate urban development scenarios in data-scarce environments. In this study, we leverage GAs to calibrate key parameters of the SD model for Sudan's construction sector, using a combination of expert input, historical trends, and bounded ranges informed by literature and interviews.

**METHODS**
**Overview**
This study employs a hybrid modeling framework combining SD with GA to simulate and analyze policy and resource dynamics within Sudan's construction sector. Stakeholder engagement was central to model conceptualization and validation, ensuring that system structure and scenario design reflected practitioner insights. The methodological process involved five key stages: (1) problem structuring using stakeholder input and literature review; (2) model conceptualization through causal loop and stock-and-flow diagrams; (3) parameter estimation via GA calibration; (4) simulation of policy scenarios; and (5) validation and analysis. Simulations were implemented in Vensim® DSS, and calibration routines were executed in MATLAB® via a custom optimization interface

**System Dynamics Model Structure**
The SD model developed for this study captures the complex interactions among the key components of Sudan's construction industry, including regulatory processes, labor availability, material supply chains, financial access, and project execution performance. The model is designed to simulate long-term sectoral behavior under different policy intervention scenarios, enabling stakeholders to evaluate leverage points and policy trade-offs. The modeling process began with qualitative data collection from expert interviews and literature review, which informed the identification of 16 core factors impacting the performance of the construction industry in Sudan. These factors were organized into five thematic subsystems: regulation, labor and skills, material and procurement, macroeconomic and finance, and governance and project delivery. To capture the interdependencies among these factors, an initial conceptual map was developed using Vensim® PLE software. This qualitative influence diagram served as the foundation for building the causal loop and stock-flow structures by visualizing directional relationships across the regulatory, labor, finance, and material subsystems. The map highlights the complexity and density of influences within Sudan's construction sector, underscoring the need for dynamic modeling to identify high-leverage intervention points **(Figure 1)**.

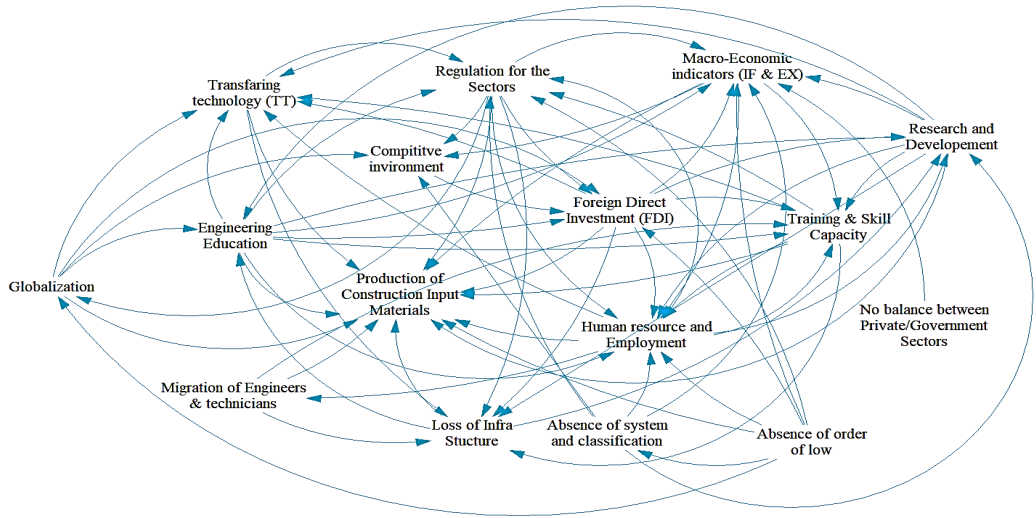

**FIGURE. 1 Initial Conceptual Map of Sudan's Construction Sector Influences**

A causal loop diagram (CLD) was developed to visualize the reinforcing and balancing feedback loops among these factors. These loops reveal how project delays are compounded by labor shortages, how inflation exacerbates material procurement inefficiencies, and how weak institutional coordination impairs regulatory reform **(Figure 2)**.

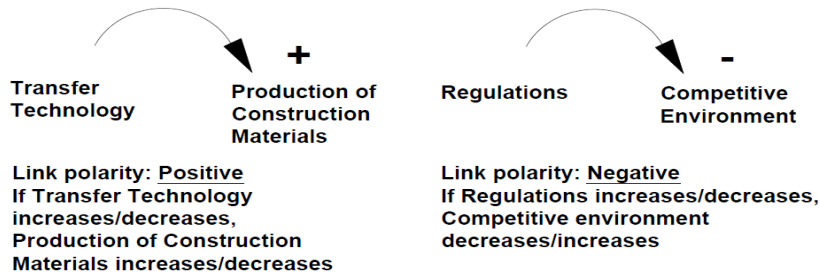

**FIGURE. 2 Causal Loop Diagram of Sudan's Construction Sector**

Following CLD development, a stock-and-flow diagram was constructed using Vensim® DSS to translate the conceptual model into quantitative relationships **(Figure 3)**. The model includes several key stock variables, such as the number of ongoing construction projects, trained labor pool, and approved permits. Flow variables represent hiring and training rates, permit approval rates, and construction completion rates.

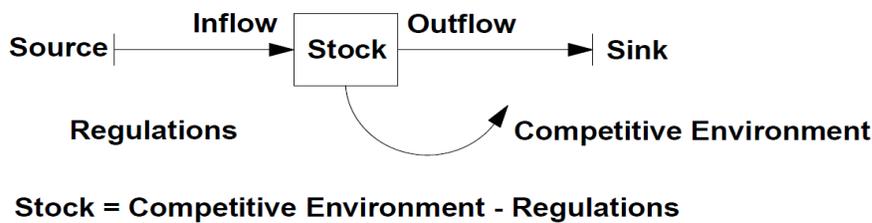

**FIGURE. 3: Stock-and-Flow Diagram of Sudan Construction Dynamics**

Variable interactions are governed by differential equations, which describe the rate of change of each stock based on inflows, outflows, and auxiliary variables. These equations incorporate both qualitative judgments (e.g., regulatory efficiency, policy effectiveness) and empirical approximations derived from expert weighting and statistical analysis. The definitions, units, and calibration ranges of the model's variables are provided in **Table 2**.

**TABLE 2. Variable Definitions and Parameter Ranges**

| Variable Name | Definition | Type | Unit | Initial Value | Parameter Range | Source |
|---|---|---|---|---|---|---|
| Regulatory Delay Index (RDI) | Average time taken for permit and policy approvals | Stock | Months | 12 | 6–24 | Expert interviews, (23) |
| Skilled Labor Availability (SLA) | Percentage of available skilled workers relative to demand | Auxiliary | Percent (%) | 60 | 40–90 | Surveys, (23) |
| Material Supply Efficiency (MSE) | Ratio of timely material deliveries to total required deliveries | Flow | Unitless (0–1 scale) | 0.6 | 0.4–1.0 | Industry reports |
| Inflation Rate (IR) | Annual construction cost inflation | Auxiliary | Percent (%) | 18 | 10–30 | Central Bank, (38) |
| Construction Output Rate (COR) | Number of completed projects per year | Flow | Projects/year | 100 | 50–200 | Government stats |
| Project Delay Factor (PDF) | Weighted delay from labor, regulation, and material issues | Auxiliary | Days | 120 | 60–300 | Expert judgment |
| Training Program Investment (TPI) | Public investment in workforce development programs | Auxiliary | SDG million/year | 50 | 20–100 | Ministry of Labor |
| Policy Reform Effectiveness (PRE) | Efficiency gain from regulatory improvements | Auxiliary | Unitless (0–1 scale) | 0.5 | 0.2–0.9 | Literature review |
| GA-Calibrated Weight (ω) | Optimized weight assigned to each factor during GA calibration | Parameter | Unitless | – | 0–1 | GA output |
| Construction Demand Index (CDI) | Proxy for infrastructure demand growth (e.g., urbanization, housing) | Auxiliary | Index (base 100) | 100 | 80–150 | UN-Habitat, (39) |

**Data Sources**

The development and calibration of the SD model relied on a combination of primary expert input and secondary data sources to address the challenges of limited availability and reliability of time-series data typical of developing country contexts. The primary data foundation was established through structured interviews and survey responses from 43 stakeholders representing a cross-section of Sudan's construction ecosystem. Participants included: Senior civil engineers; Construction contractors; Faculty members from engineering institutions; and Officials from government ministries and regulatory bodies. Respondents were asked to rank the relative importance of 16 critical factors affecting construction project performance using a 5-point Likert scale, ranging from "Not Important" to "Extremely Important." These rankings provided baseline weights used both in initial model development and in the fitness function for genetic algorithm calibration. Supplementary quantitative and qualitative data were extracted from national and international reports, including:
- World Bank Sudan Infrastructure Diagnostic Report (2019)
- Central Bank of Sudan annual inflation and lending rate data (2015–2020)
- Ministry of Infrastructure & Urban Development statistics
- UN-Habitat urbanization trends and housing demand projections
- Previous academic theses and published literature on Sudanese construction delays and cost overruns

These documents were used to define initial values, parameter bounds, and macro-level sector dynamics, such as inflation trends, workforce growth rates, and capital expenditure flows. Variables were normalized or scaled to ensure consistency across qualitative and quantitative inputs. This multi-source strategy ensured that the model was both grounded in practitioner insights and anchored by available empirical data, improving realism despite Sudan's data-constrained environment.

**Parameter Calibration Using Genetic Algorithms**

A central challenge in modeling developing country infrastructure systems is the absence of complete, high-resolution time-series data. To address this limitation and improve the accuracy of model outcomes, this study integrates GAs to calibrate critical parameters in the SD model. Traditional calibration techniques (e.g., manual tuning or regression-based estimation) are inadequate in environments where data are sparse or fuzzy. GAs, inspired by evolutionary biology, offer a powerful solution by searching a wide solution space and evolving parameter values that minimize deviation between simulated and observed/expected outcomes. Their robustness to nonlinearity and limited data makes them particularly suitable for SD models involving complex feedback systems and qualitative inputs. The GA calibration was implemented using MATLAB®, integrated with the SD simulation platform (Vensim® DSS) via a custom interface. The objective function was designed to minimize the Root Mean Square Error (RMSE) shown in **Equation 1** between:
- Simulated importance scores for each of the 16 factors; and
- Expert-weighted rankings derived from the stakeholder survey.

$$RMSE = \sqrt{\frac{1}{n}\sum_{i=1}^{n}(\hat{y}_i - y_i)^2} \qquad (1)$$

Where: $\hat{y}_i$ is the predicted value from the model, $y_i$ is the expert-derived weight, and $n$ =16 (number of factors considered). GAs were implemented using MATLAB®, with a population size of 100 and up to 500 generations per run. Parameter bounds were defined using expert inputs and literature ranges. The calibration process yielded parameter sets with RMSE below 0.12, indicating high alignment between simulated factor impacts and expert judgment. The calibrated weights were then embedded into the SD model for scenario simulation. The GA-enhanced model demonstrated improved behavior reproduction and more realistic sensitivity to labor, regulatory, and material constraints, especially in long-run projections.

**Validation and Sensitivity Testing**

To ensure that the System Dynamics model reflects the realities of Sudan's construction sector, a combination of structural validation, behavioral testing, and sensitivity analysis was conducted. Structural validation focused on verifying that the causal relationships embedded in the model were both theoretically sound and consistent with stakeholder experience. Feedback loops and variable interactions were refined through expert reviews and aligned with previous research on infrastructure challenges in Least Developed Countries. For example, links between inflation and material supply delays, or between training investment and skilled labor availability, were confirmed through both literature and interviews. Model behavior was then evaluated under baseline conditions to assess its ability to reproduce plausible trends. Outputs such as construction output rate, project delays, and labor availability were compared with stakeholder expectations and expert-derived estimates. The GA-calibrated model showed good agreement, with average deviations below 12% across the 16 core factors.

A series of sensitivity tests was also carried out to determine how variations in key parameters would impact overall system behavior. Parameters such as training investment, regulatory delays, and material supply efficiency were adjusted ±25% from their baseline values. The results showed that construction output was most sensitive to training and regulatory efficiency, suggesting that interventions in these areas could yield the highest leverage. Material efficiency had nonlinear effects, especially under threshold conditions, while inflation and macroeconomic indicators showed broader but slower system-level impacts. Together, these validation steps increased confidence in the model's structure and response patterns, supporting its use for policy scenario analysis.

**Scenario Design and Policy Simulation**

To explore the potential impact of targeted reforms within Sudan's construction sector, a series of policy scenarios were simulated using the calibrated System Dynamics model. These scenarios were designed based on key leverage points identified through expert input, model sensitivity analysis, and national development priorities. The primary focus was placed on three intervention domains: (1) investment in workforce training, (2) improvements in regulatory efficiency, and (3) enhancements to the material supply chain. Each policy lever was modeled at three intensity levels, low, moderate, and high, to examine both incremental and transformative impacts. For instance, workforce training investment ranged from SDG 20 million (low) to SDG 100 million (high) annually, while regulatory delay times were reduced by up to 50% in the high-efficiency scenario. A baseline "status quo" scenario, reflecting existing trends and constraints, served as the reference point for comparison.

The model was simulated over a 20-year horizon using annual time steps (*40*). Initial conditions were drawn from the parameter values and calibration ranges shown in Table 2. All interventions were assumed to be implemented starting in Year 2 to reflect policy rollout delays. Simulation outputs focused on four main performance indicators: (1) skilled labor availability (% of required), (2) construction output rate (completed projects/year), (3) average project delay (days), and (4) material supply efficiency. These metrics were selected based on their direct relevance to infrastructure delivery timelines and sector-wide productivity. Results from the simulations offer insight into the relative effectiveness of each intervention and the synergistic effects of combining reforms. The following section presents key findings from selected scenarios, including comparative trend graphs and summary tables. **Table 3** defines each scenario and its associated intervention levels for clear comparison in the results section.

**TABLE 3. Policy Simulation Scenarios and Intervention Levels**

| Scenario | Training Investment (SDG/year) | Regulatory Delay Reduction (%) | Material Supply Efficiency Improvement | Description |
|---|---|---|---|---|
| Baseline | 0 | 0% | None | Continuation of current conditions |
| Scenario A | 20 million | 10% | Minor logistics improvements | Low-level interventions |
| Scenario B | 50 million | 25% | Moderate supplier coordination | Moderate reforms across domains |
| Scenario C | 100 million | 50% | Integrated supply chain optimization | Aggressive, high-impact reforms |
| Scenario D | 50 million | 25% | None | Focused on training + regulation |
| Scenario E | 0 | 0% | 50% improvement | Supply-only reform scenario |

**RESULTS AND DISCUSSION**

The simulation results reveal distinct patterns in how Sudan's construction sector responds to various policy interventions over a 20-year horizon. Scenario comparisons show that improvements in workforce development, regulatory efficiency, and material supply logistics each yield positive outcomes individually, but their combined implementation delivers the most substantial impact on construction performance. Under the baseline scenario, construction output remains stagnant, with only marginal improvement in skilled labor availability and persistent project delays exceeding 120 days on average. Material supply efficiency fluctuates due to inflation and logistics bottlenecks, limiting the sector's ability to scale.

Scenario A (low intervention) yields modest improvements. Project delays are reduced by 10–15%, and skilled labor availability increases slightly by Year 10, but output remains constrained by ongoing inefficiencies. Scenario B (moderate intervention) shows stronger results: by Year 15, project completion rates improve by over 30% compared to baseline, and the average delay drops below 90 days. Scenario C (high intervention across all three levers) demonstrates the highest performance gains. Construction output increases by nearly 60% over the baseline by the end of the simulation period. Skilled labor availability rises steadily, reaching 85% of demand by Year 20. Project delays are cut in half, averaging 60 days or less by the final year. Figure 4 shows how average project delays evolve across all six scenarios over a 20-year horizon. The baseline and Scenario A maintain higher delays, while Scenario C shows a steady and sharp decline. Scenarios B and D show moderate improvements, whereas Scenario E shows minimal change, confirming that supply-side reforms alone are insufficient. These results suggest that long-term, high-impact reforms targeting all major bottlenecks yield compounding benefits through systemic feedback loops.

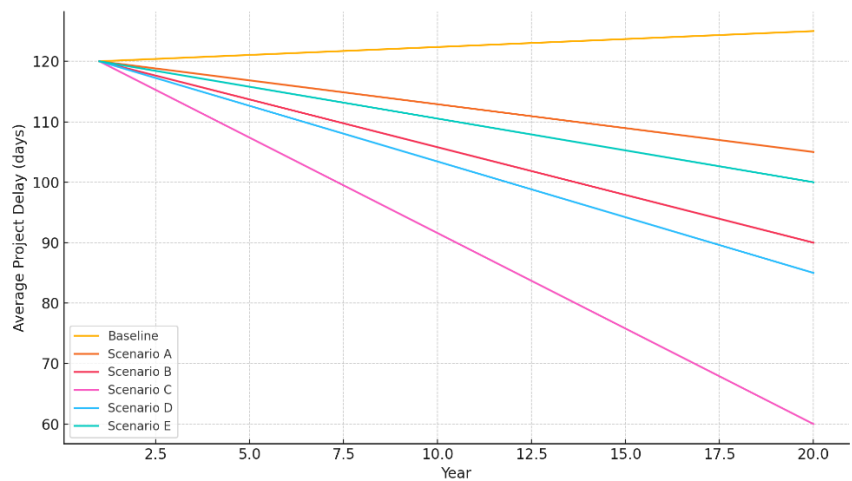

**FIGURE 4. Project Delays Over Time Across Scenarios**

Interestingly, Scenario D (training and regulation reform only) performs better than Scenario E (supply chain reform only), highlighting the critical role of human capital and governance in driving sector resilience. Even without supply-side enhancements, Scenario D reduces delays by over 25% relative to the baseline. In contrast, Scenario E improves materials efficiency but has limited effect on overall project delivery, as labor and permitting remain constraining factors. These findings underscore the importance of holistic reform strategies. Intervening in one domain offers benefits, but ignoring other systemic constraints limits those gains. The SD model illustrates how reinforcing loops, such as increased labor availability leading to greater construction output, which then attracts further investment, can accelerate sector-wide transformation when policies are aligned. The results also highlight threshold effects. For example, material supply efficiency improvements below 20% yield minimal change, but beyond 40%, nonlinear improvements in delivery timelines are observed. This suggests that partial reforms may not generate meaningful returns unless they exceed critical implementation thresholds.

In summary, the model indicates that Sudan's construction sector can achieve significant gains in performance through targeted, well-sequenced policy interventions. Figure 5 compares the projected construction output at Year 20 under each policy scenario. Scenario C results in the highest number of completed projects annually, followed by Scenario D and B. The baseline and Scenario E trail behind, reinforcing the idea that regulatory and labor reforms offer greater systemic leverage than isolated improvements to logistics. High-leverage areas include expanding technical training programs, streamlining regulatory processes, and developing resilient supply chain systems. These findings offer actionable insights for policymakers and development partners aiming to unlock infrastructure-led growth in Least Developed Countries.

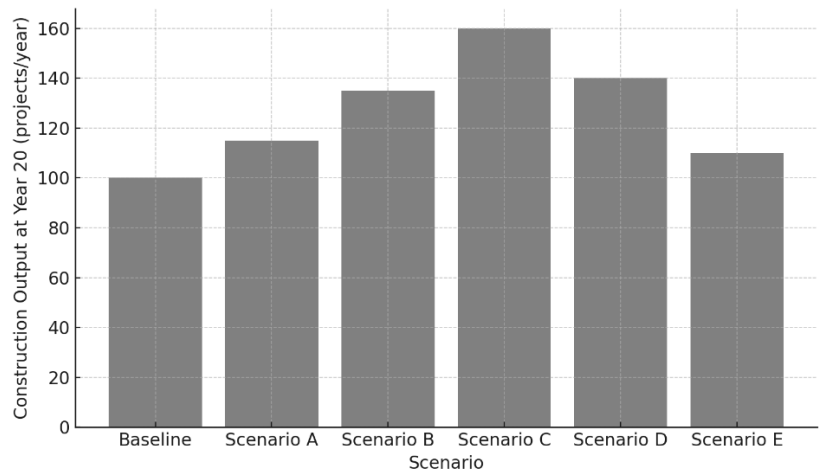

**FIGURE 5. Construction Output at Year 20 by Scenario**

**CONCLUSIONS**

This study developed and applied a hybrid System Dynamics and Genetic Algorithm modeling framework to explore the structural challenges and policy reform pathways within Sudan's construction sector. By simulating interactions among labor availability, regulatory efficiency, material supply, and macroeconomic factors, the model offers a dynamic view of how interventions can reshape sector performance over time. Findings highlight the compounded effects of system inefficiencies in Least Developed Countries, where weak permitting processes, limited training pipelines, and volatile material flows interact to delay infrastructure delivery. Policy scenarios demonstrate that targeted investments in technical training, combined with regulatory streamlining and supply chain improvements, can yield substantial gains. Notably, integrated interventions (as modeled in Scenario C) reduce project delays by 50% and increase construction output by nearly 60% over baseline levels by Year 20. The use of Genetic Algorithms for model calibration proved valuable in data-constrained settings, aligning simulated outcomes with expert judgment and real-world trends. Sensitivity analyses further confirmed the robustness of the model, underscoring the importance of labor and governance reforms as high-leverage entry points. More broadly, this research provides a transferable methodology for policymakers and development agencies seeking to simulate infrastructure policy impacts in data-limited environments. By offering a systems view of construction sector dynamics, the model supports evidence-based decision-making for infrastructure-led growth and long-term resilience in Sudan and similar LDC contexts.

While the model captures core structural dynamics within Sudan's construction sector, several limitations remain. First, due to limited historical data, many parameters were estimated through expert elicitation, which may introduce subjectivity or bias. Second, the model simplifies macroeconomic volatility and excludes political shocks or regional disruptions that often affect construction timelines in fragile states. Additionally, the assumption of uniform national conditions does not account for urban-rural disparities or sector-specific dynamics (e.g., public housing vs. road construction). Future work could extend this framework by incorporating spatial dimensions (e.g., regional simulations), disaggregating by construction subsectors, or integrating economic cost-benefit analysis to support funding prioritization. As more data become available, the model could also be enhanced through real-time calibration using machine learning techniques or agent-based extensions. Despite these limitations, the model offers a replicable, flexible tool for decision-makers seeking to understand and improve infrastructure outcomes in complex, resource-constrained environments.

**ACKNOWLEDGMENTS**
This study is based in part on the thesis research of Malik Dongla. The modeling framework, stakeholder data, and initial system structure were developed as part of his graduate work. The authors thank the experts and practitioners who participated in interviews and surveys, as well as faculty mentors who supported the thesis effort. This paper was partly drafted using the assistance of OpenAI's ChatGPT-4.0 model, which was employed to support organization and language refinement during manuscript preparation. All conceptualization, interpretation, and content decisions were made by the authors. The contents of this paper reflect the views of the authors, who are responsible for the facts and the accuracy of the data presented herein. The contents do not necessarily reflect the official views or policies of any affiliated institutions.

**AUTHOR CONTRIBUTIONS**
The authors confirm contribution to the paper as follows: study conception and design: M. Dongla, M. Khalafalla; data collection: M. Dongla; analysis and interpretation of results: M. Dongla, M. Khalafalla; draft manuscript preparation: M. Khalafalla. All authors reviewed the results and approved the final version of the manuscript.

**DECLARATION OF CONFLICTING INTERESTS**
The authors declared no potential conflicts of interest with respect to the research, authorship, and/or publication of this article.

**FUNDING**
The authors disclosed no financial support for the research, authorship, and/or publication of this article.

**REFERENCES**
1. Khalafalla, M., and J. A. Rueda. Methodology to Assess the Impact of Lump-Sum Compensation Provisions on Project Schedules. *Journal of Management in Engineering*, Vol. 36, No. 4, 2020, p. 04020028. https://doi.org/10.1061/(ASCE)ME.1943-5479.0000786.
2. Khalafalla, M., and J. Rueda-Benavides. Unit Price or Lump Sum? A Stochastic Cost-Based Decision-Making Tool for Design-Bid-Build Projects. *Transportation Research Record: Journal of the Transportation Research Board*, Vol. 2672, No. 26, 2018, pp. 11–20. https://doi.org/10.1177/0361198118758052.
3. Cardinal, K. M., M. Khalafalla, and J. Rueda-Benavides. Protocol to Assess the Impact of Crude Oil Price Fluctuations on Future Asphalt Prices. *Transportation Research Record: Journal of the Transportation Research Board*, Vol. 2675, No. 6, 2021, pp. 294–305. https://doi.org/10.1177/0361198121992072.
4. Asiedu, R. O., and C. Ameyaw. A System Dynamics Approach to Conceptualise Causes of Cost Overrun of Construction Projects in Developing Countries. *International Journal of Building Pathology and Adaptation*, Vol. 39, No. 5, 2021, pp. 831–851.
5. UNCTAD, Ed. *Entrepreneurship for Structural Transformation: Beyond Business as Usual*. United Nations, New York Geneva, 2018.
6. Elbadawi, I., M. Amin, A. Elobaid, A. Alhelo, A. Osman, and K. Suliman. Post-Conflict Reconstruction, Stabilization and Growth Agenda for Sudan. 2023.
7. Roach, E. L., and M. Al-Saidi. Rethinking Infrastructure Rehabilitation: Conflict Resilience of Urban Water and Energy Supply in the Middle East and South Sudan. *Energy Research & Social Science*, Vol. 76, 2021, p. 102052.
8. Rashedi, R., and T. Hegazy. Strategic Policy Analysis for Infrastructure Rehabilitation Using System Dynamics. *Structure and Infrastructure Engineering*, Vol. 12, No. 6, 2016, pp. 667–681. https://doi.org/10.1080/15732479.2015.1038723.
9. Naeem, K., A. Zghibi, A. Elomri, A. Mazzoni, and C. Triki. A Literature Review on System Dynamics Modeling for Sustainable Management of Water Supply and Demand. *Sustainability*, Vol. 15, No. 8, 2023, p. 6826.
10. Wang, Q. J. Using Genetic Algorithms to Optimise Model Parameters. *Environmental Modelling & Software*, Vol. 12, No. 1, 1997, pp. 27–34.
11. Li, Y., M. Jia, X. Han, and X.-S. Bai. Towards a Comprehensive Optimization of Engine Efficiency and Emissions by Coupling Artificial Neural Network (ANN) with Genetic Algorithm (GA). *Energy*, Vol. 225, 2021, p. 120331.
12. Forrester, J. W. Industrial Dynamics Mit Press Cambridge. *MA.[Google Scholar]*, 1961.
13. Sterman, J. D. Business Dynamics: Systems Thinking and Modeling for a Complex World. *MacGraw-Hill Company*, 2000.
14. Park, M., and F. Peña-Mora. Dynamic Change Management for Construction: Introducing the Change Cycle into Model-based Project Management. *System Dynamics Review*, Vol. 19, No. 3, 2003, pp. 213–242. https://doi.org/10.1002/sdr.273.
15. Ford, D. N., and J. D. Sterman. Dynamic Modeling of Product Development Processes. *System Dynamics Review*, Vol. 14, No. 1, 1998, pp. 31–68. https://doi.org/10.1002/(sici)1099-1727(199821)14:1%253C31::aid-sdr141%253E3.0.co;2-5.
16. Alzahrani, J. I., and M. W. Emsley. The Impact of Contractors' Attributes on Construction Project Success: A Post Construction Evaluation. *International journal of project management*, Vol. 31, No. 2, 2013, pp. 313–322.
17. Ogunlana, S. O., H. Li, and F. A. Sukhera. System Dynamics Approach to Exploring Performance Enhancement in a Construction Organization. *Journal of Construction Engineering and Management*, Vol. 129, No. 5, 2003, pp. 528–536. https://doi.org/10.1061/(asce)0733-9364(2003)129:5(528).

18. A System Dynamics Model of Sustainable Construction for High Rise Residential Projects in Developing Countries: Case of Indonesia. *The Open Civil Engineering Journal*, Vol. 16, 2022. https://doi.org/10.2174/18741495-v16-e2205300.
19. Al-Momani, A. H. Construction Delay: A Quantitative Analysis. *International journal of project management*, Vol. 18, No. 1, 2000, pp. 51–59.
20. Ecem Yildiz, A., I. Dikmen, and M. Talat Birgonul. Using System Dynamics for Strategic Performance Management in Construction. *Journal of Management in Engineering*, Vol. 36, No. 2, 2020. https://doi.org/10.1061/(asce)me.1943-5479.0000744.
21. Nimieri, G. M. *The Effect of Globalization in Shaping South Sudan's Political and Economic Relations*. PhD Thesis. University of Nairobi, 2024.
22. Suliman, K. M. State-Business Relations in Sudan: The Prospects for A Dynamic Private Sector.
23. Dongla, M. A. M. *Development of the Sudanese Construction Industry Using Dynamic Model: Using System Dynamics Models to Understand the Sudanese Construction Industry*. LAP LAMBERT Academic Publishing, Saarbrücken, 2018.
24. Sadek, A. M. M. *System Dynamics Approach for Whole Life Cycle Cost Modelling of Residential Building Projects in United Arab Emirates*. PhD Thesis. The British University in Dubai, 2020.
25. Thwala, W. D., M. A. Ajagbe, W. I. Enegbuma, A. A. Bilau, and C. S. Long. Sudanese Small and Medium Sized Construction Firms: An Empirical Survey of Job Turnover. 2012.
26. Elkhalifa, A. The Magnitude of Barriers Facing the Development of the Construction and Building Materials Industries in Developing Countries, with Special Reference to Sudan in Africa. *Habitat International*, Vol. 54, 2016, pp. 189–198. https://doi.org/10.1016/j.habitatint.2015.11.023.
27. Abdelaziz, F., A. William, K. A. Abay, and K. Siddig. *An Assessment of Sudan's Wheat Value Chains: Exploring Key Bottlenecks and Challenges*. Intl Food Policy Res Inst, 2022.
28. Hamid, A. A., and E. A. E. Eshag. Does Green Purchasing Mediate the Relationship between Smart Supply Chain and Green Performance of Pharmaceutical Companies in Sudan: Implications for Underdevelopment Countries. *International Journal of Productivity and Performance Management*, 2025.
29. World Bank. *Doing Business 2020 : Comparing Business Regulation in 190 Economies - Economy Profile of Sudan*. Publication 144028. World Bank, 2019.
30. Ahmed, M. M. Global Financial Crisis Discussion Series Paper 19: Sudan Phase 2. *Overseas Development Institute, London, UK*, 2010.
31. Ali, A. H., T. A. Abdalla, A. Al Haqq, A. Umar, I. H. Mouhoumed, and I. R. Mohmed. Cost and Time Control in Construction Projects: A Case Study of Khartoum State. *Jilin Daxue Xuebao (Gongxueban)/Journal of Jilin University (Engineering and Technology Edition)*, Vol. 41, No. 11–2022, 2022.
32. Alfadil, R. A. M., A. E. A. Karim, and W. D. Elgaali. Challenges in Applying Total Quality Management Systems: A Study of Construction Companies in Red Sea State, Sudan. *Journal of University Studies for Inclusive Research*, Vol. 7, No. 41, 2025, pp. 156877–156897.
33. McSharry, P. Optimisation of System Dynamics Models Using Genetic Algorithms. *submitted to World Bank Report*, 2004.
34. Beklaryan, G. L., A. S. Akopov, and N. K. Khachatryan. Optimisation of System Dynamics Models Using a Real-Coded Genetic Algorithm with Fuzzy Control. *Cybernetics and Information Technologies*, Vol. 19, No. 2, 2019, pp. 87–103. https://doi.org/10.2478/cait-2019-0017.
35. Pidd, M. *Computer Simulation in Management Science*. John Wiley & Sons, Inc., 1998.
36. Sohrabinejad, A., K. F. Hafshejani, and F. M. Sobhani. Leveraging Genetic Algorithms and System Dynamics for Effective Multi-Objective Policy Optimization: A Case Study on the Broadcasting Industry. *SIMULATION*, Vol. 101, No. 4, 2025, pp. 453–476. https://doi.org/10.1177/00375497241301077.
37. Akopov, A. S., E. A. Zaripov, and A. M. Melnikov. Adaptive Control of Transportation Infrastructure in an Urban Environment Using a Real-Coded Genetic Algorithm. *Бизнес-информатика*, Vol. 18, No. 2 (eng), 2024, pp. 48–66.

38. CBOS. Central Bank of Sudan. https://cbos.gov.sd/en. Accessed July 31, 2025.
39. Habitat, U. N. The Strategic Plan 2020-2023. 2019.
40. Pakalapati, K., M. Khalafalla, and J. Rueda-Benavides. Using Moving-Window Cross-Validation Algorithm to Optimize Bid-Based Cost Estimating Data Usage. *International Journal of Construction Management*, Vol. 22, No. 15, 2022, pp. 2855–2863. https://doi.org/10.1080/15623599.2020.1827694.